\documentclass[a4paper,11pt]{article}
\usepackage{graphicx}
\usepackage{fullpage}

\usepackage{jheppub} 

\usepackage{booktabs}
\usepackage{subcaption}
\usepackage{framed}
\usepackage{multirow}
\usepackage{array}
\usepackage{lscape}
\usepackage{rotating}
\usepackage{float}

\def\be{\begin{equation}}
\def\ee{\end{equation}}
\def\bea{\begin{array}}
\def\eea{\end{array}}
\def\beqa{\begin{eqnarray}}
\def\eeqa{\end{eqnarray}}
\def\beqas{\begin{eqnarray*}}
\def\eeqas{\end{eqnarray*}}

\def\bp{\begin{picture}}
\def\ep{\end{picture}}
\def\bc{\begin{center}}
\def\ec{\end{center}}
\def\bfig{\begin{figure}}
\def\efig{\end{figure}}

\def\bit{\begin{itemize}}
\def\eit{\end{itemize}}
\def\nn{\nonumber}
\def\f{\frac}

\def\[{\left[}
\def\]{\right]}
\def\({\left(}
\def\){\right)}

\def\..{\left.}
\def\.{\right.}
\def\tl{\tilde}
\def\ra{\rightarrow}
\def\la{\leftarrow}

\def\tm{\times}

\def\NPB#1,{{ Nucl.\ Phys.\ B }{\bf #1},}
\def\PLB#1,{{ Phys.\ Lett.\ B }{\bf #1},}
\def\EPJC#1,{{ Eur.\ Phys.\ Jour.\ C }{\bf #1},}
\def\PRD#1,{{ Phys.\ Rev.\ D }{\bf #1},}
\def\PRL#1,{{ Phys.\ Rev.\ Lett.\ }{\bf #1},}
\def\MPLA#1,{{Mod.\ Phys.\ Lett.\ A }{\bf #1},}

\def\da{\dagger}

\def\la{\lambda}

\def\al{\alpha}

\def\ep{\epsilon}

\def\pa{\partial}

\title{\sffamily Heavy colored SUSY partners from deflected anomaly mediation}

\author[a,b]{Fei Wang,}
\author[c]{Wenyu Wang,}
\author[b]{Jin Min Yang,}
\author[b]{Yang Zhang}

\affiliation[a]{Department of Physics and Engineering, Zhengzhou University, Zhengzhou 450000,
                P. R. China}
\affiliation[b]{State Key Laboratory of Theoretical Physics, Institute of Theoretical Physics,
                Academia Sinica, Beijing 100190, P. R. China}
\affiliation[c]{Institute of Theoretical Physics, College of Applied Science,
Beijing University of Technology, Beijing 100124, P. R. China}

\emailAdd{feiwang@zzu.edu.cn}
\emailAdd{wywang@mail.itp.ac.cn}
\emailAdd{jmyang@itp.ac.cn}
\emailAdd{zhangyang@itp.ac.cn}

\abstract{We propose a deflected anomaly mediation scenario
from SUSY QCD which can lead to both positive and negative deflection parameters
(there is a smooth transition between these two deflection parameter
regions by adjusting certain couplings). Such a scenario
can naturally give a SUSY spectrum in which all the colored sparticles are heavy
while the sleptons are light. As a result, the discrepancy between the Brookheaven $g_\mu-2$
experiment and LHC data can be reconciled in this scenario.
We also find that the parameter space for explaining the $g_\mu-2$ anomaly at $1\sigma$ level
can be fully covered by the future LUX-ZEPLIN 7.2 Ton experiment.
}

\begin{document}
\maketitle \indent
\newpage

\section{Introduction}
As an appealing candidate for the TeV-scale new physics,
low energy supersymmetry (SUSY) can give an explanation for the gauge hierarchy
problem, realize the gauge coupling unification and provide a viable dark matter
candidate. It is remarkable that the 125 GeV Higgs boson recently discovered by
the ATLAS \cite{ATLAS:higgs} and CMS collaborations \cite{CMS:higgs}
agrees perfectly with the mass prediction of 115-135 GeV by the Minimal
Supersymmetric Standard Model (MSSM).
Actually, SUSY can satisfy all current experimental constraints \cite{cao}
and especially can yield sizable contributions
to the muon anomalous magnetic moment which can solve the discrepancy
between the E821 experiment at the Brookhaven AGS \cite{BK:g-2,BK2:g-2}
and the Standard Model (SM) prediction \cite{SM:g-2}.

On the other hand, so far no SUSY partners have been detected at the LHC
and the mass limits on squarks and gluinos $m_{\tilde g} > 1.5$ TeV for
$m_{\tl{q}} \sim m_{\tl{g}}$ and $m_{\tl{g}}\gtrsim 1$ TeV for $m_{\tl{q}} \gg m_{\tl{g}}$
have been obtained for the constrained MSSM (CMSSM) \cite{CMSSM1,CMSSM2}.
Together with the heavy top squarks required by the 125 GeV Higgs boson mass
(the mass bounds from the direct LHC search are not so stringent for top squarks \cite{han}),
this indicates rather heavy colored sparticles \footnote{However, the recent ATLAS Z-peaked
excess \cite{ATLAS-Z-Excess} may indicate a gluino as light as 800 GeV \cite{Explanation-susy}}.
Considering the light uncolored sparticles (neutralinos, charginos and
smuons) around ${\cal O}(100)$ GeV required by the explanation of the muon
$g_\mu-2$ anomaly \cite{Mg-2:SUSY}, this poses a tension for the popular
CMSSM \cite{cao-cmssm,Nath}. So the SUSY spectrum from SUSY breaking seems to have a intricate
structure \cite{ewsusy}.
The origin of SUSY breaking and its mediation mechanism
are crucial for the phenomenology.

The anomaly mediated SUSY breaking (AMSB) \cite{AMSB} is one of the
most attractive scenarios in supergravity. Not only the sparticle mass spectrum
are predicted to be flavor blind and thus automatically solves the SUSY flavor problem,
but also the sparticle masses at low energies are insensitive to any high energy
theories \cite{deflect:RGE-invariance} since the SUSY breaking is mediated through
the superconformal anomaly. Unfortunately, the AMSB scenario leads to tachyonic
sleptons so that the minimal theory must be extended. The deflected AMSB \cite{deflect},
which introduces a messenger sector in the AMSB, can deflect the Renormalization
Group Equation (RGE) trajectory and give new contributions to the soft SUSY breaking terms.
The tachyonic slepton problems can be naturally solved by such a deflection.

The SUSY spectrum with heavy colored sparticles and light sleptons
can be naturally realized in such a deflected AMSB scenario, especially when
the deflection parameter is positive \cite{okada,carpenter}.
However, the positive-deflected AMSB model cannot be easily realized and special efforts
are need for model building. We propose in this paper a scenario from SUSY QCD
in which both positive and negative deflection parameters can be realized and smoothly
connected. Messenger sectors can be generated automatically without additional assumptions.
With positive-deflected parameters, the tension between $g_\mu-2$ anomaly and
LHC data can be ameliorated in such a AMSB scenario.

This paper is organized as follows.  In Sec. \ref{review}, we briefly review
the AMSB mechanism. In Sec.\ref{model}, we propose a
scenario which can realize both positive and negative deflected AMSB from SUSY QCD-type theory.
A smooth transition can occur for both possibilities and all the contents can origin
from a SUSY strong dynamics. In Sec. \ref{phenomenology}, we examine the parameter space
of our deflected AMSB to explain both the LHC results and the Brookhaven $g_\mu-2$ experiments.
Sec. \ref{conclusion} contains our conclusions.

\section{A brief review of AMSB} \label{review}
In the MSSM, the SUSY breaking effects can be communicated from some
hidden sector to visible sector through gauge \cite{GMSB} or
gravitational \cite{SUGRA} interactions. Gravitational effects typically lead
to sparticle masses from contact terms suppressed by powers of the Planck scale.
However, if the two sectors are completely sequestered and these contact terms are absent,
the sparticle masses of order $m_{3/2}/(16\pi^2)$ will still be generated due to the
superconformal anomaly \footnote{The analysis in \cite{Seiberg:Dine} clarifies
several physical aspects of AMSB and demonstrates that anomaly mediation
of SUSY breaking is in fact not a consequence of any anomaly of the theory \cite{anomaly:fake}.}.
Anomaly mediation can be regarded as the pure supergravity contributions to the  soft SUSY breaking terms.
They are determined by the VEV of the auxiliary compensator field $F_\phi$ within the graviton
supermultiplet.
The couplings of compensator F-term VEVs to the MSSM are purely quantum effects from the super-Weyl anomaly.
The supergravity effects can be studied in the superconformal tensor calculus formalism
by the introduction of compensator field \cite{compensator} . The theory with the compensator can be seen to
be equivalent to ordinary non-conformal SUGRA after gauge fixing.

Assume that the only source of SUSY breaking comes from a non-vanishing value of $F_\phi$ of compensator
field with
\beqa
\langle\phi\rangle=1+\theta^2 F_\phi.
\eeqa
and $F_\phi\sim m_{3/2}$.
The couplings of $\phi$ are restricted by a spurion scale symmetry under which $\phi$
has a mass dimension of +1. Therefore, $\phi$ only appears in terms with dimensionful
couplings. Although there is no SUSY breaking at tree level, soft masses
at loop level emerges because the cut-off in a supersymmetric regulator
is a dimensionful coupling which must be made covariant.
The Lagrangian of the visible sector can be written as
\beqa
{\cal L}=\int d^4\theta Z\left(\f{\mu}{\Lambda}\right)Q^\da e^V Q
 +\int d^2\theta \left[S\left(\f{\mu}{\Lambda}\right)W^aW_a+\la Q^3\right]+h.c.
\eeqa
After the replacement
\beqa
 Z\left(\f{\mu}{\Lambda}\right)\ra  Z\left(\f{\mu}{\Lambda\sqrt{\phi^\da\phi}}\right)~,
~~~~S\left(\f{\mu}{\Lambda}\right) \ra S\left(\f{\mu}{\Lambda\sqrt{\phi^\da\phi}}\right),
\eeqa
and the expansion in $\theta$, the gauginos acquire masses
\beqa
M_\la=\f{g^2}{2}\f{d g^{-2}}{d \ln\mu} F_\phi=\f{b g^2}{16\pi^2}F_\phi~,
\eeqa
which are typically at the scale $m_{3/2}$.
The sfermions, on the other hand, acquire masses of the form
\beqa
M_{\tl{f}}^2=-\f{1}{4}\left(\f{\pa \gamma}{\pa g}\beta_g+\f{\pa \gamma}{\pa y} \beta_y\right)\left|F_\phi\right|^2~,
\eeqa
where at leading order
\beqa
\beta_g=-\f{ b g^3}{16\pi^2},~~\gamma=\f{1}{16\pi^2}\(4 C_2(r) g^2-a_1 y^2\),~~
\beta_y=\f{y}{16\pi^2}(a_2 y^2-a_3 g^2).
\eeqa
So they give
\beqa
M_{\tl{f}}^2=\f{1}{512\pi^4}\[4 C_2(r) b g^4+a_1y^2(a_2 y^2-a_3 g^2)\]\left|F_\phi\right|^2.
\eeqa
Sfermion masses are in practice family independent but the squared slepton masses
are predicted to be tachyonic in this minimal scenario.

\section{Deflected AMSB from SUSY QCD}\label{model}
Various attempts have been proposed to solve the tachyonic slepton problem for the AMSB.
For example, additional gravitational contributions or additional D-term contributions
\cite{AMSB:D-term,deflect:RGE}
can be added to overcome this problem. It is also possible to generate large
Yukawa couplings for sleptons with additional Higgs doublets \cite{AMSB:yukawa}.
An elegant solution is the `deflected anomaly mediation' scenario \cite{deflect}
in which the soft spectrum is modified by the presence of a light modulus (massless in the
supersymmetric limit). Such a negatively deflected anomaly
mediation scenario tends to release the gaugino hierarchy at the electroweak scale and drag
down some of the squark masses, which is not favored by
the null search results of sparticles at the LHC.
In order to have relatively heavy colored sparticles and light sleptons,
we need a positively deflected scenario \cite{okada}
which gives a possible realization with some particular choice of the power for the singlet $S$.
In our following analysis, we will show that such a positively deflected scenario can be fairly generic
in SUSY QCD type theory. The messenger fields, including their couplings, are also naturally obtained
from the SUSY QCD dynamics.

We start from a microscopic model of $SU(N)$ SUSY QCD  with $N_F$ flavor, where we require
$N+1<N_F<3N$ so that the theory is asymptotic free in the UV limit and confines at the
scale $\Lambda$.
The global symmetry of the theory is $SU(N_F)_L\times SU(N_F)_R\times U(1)_V\times U(1)_R$.
We can weakly gauge the subgroup of the global symmetry to accommodate the standard model gauge group.
In terms of $SU(N)\times SU(N_F)_L\times SU(N_F)_R\times U(1)_V\times U(1)_R$ group,
the quantum number of matter contents $Q_i,\bar{Q}_j$ are given as
 \beqa
 Q_i\sim(N, N_F,~1,~1, (N_F-N)/N_F),~~~\tl{Q}_j\sim (\bar{N},~1, \bar{N}_F,-1, (N_F-N)/N_F)~.
 \eeqa
 The superpotential in the microscopic '{\em electric}' description is introduced as the ISS-type \cite{ISS}
 \beqa
 W= {\rm Tr}(m_0 \tl{Q}_i Q_i)~,
 \eeqa
which below the confining scale $\Lambda$ will have an alternative '{\em magnetic}'
description in terms of $SU(N_F-N)$ gauge theory via Seiberg duality with the following superpotential
\beqa
W=-h\tl{\mu}^2 {\rm Tr} \Phi + h  {\rm Tr} {q} \Phi \tl{q}~,
\eeqa
with $q,\tl{q}$ and $\Phi$ related to the dual baryon $B$ and meson $M$, respectively.
 The parameters are defined as
\beqa
&&\Phi=\f{M}{\sqrt{\al}\Lambda}~,~h=\f{\sqrt{\al}\Lambda}{\hat{\Lambda}}~,
~\tl{\mu}^2=-m_0\hat{\Lambda}~,~\Lambda_m=\tl{\Lambda},~N_c=N_F-N,~\nn\\
&& \Lambda^{3N-N_F}\Lambda_m^{3N_c-N_F}=(-1)^{N_c}\hat{\Lambda}^{N_F}~,
\eeqa
where $\al$ which determines the coupling $h$ is a dimensionless parameter in the Kahler potential for $M$.

It is well known that SUSY QCD of vector type does not break SUSY \cite{Witten}.
So this theory leads to a metastable SUSY breaking vacua \cite{ISS} and at the same
time has SUSY preserving vacua at large field value. In our scenario, SUSY breaking
arises from the anomaly mediation effects instead of the ISS-type rank conditions.
So we will concentrate on the originally SUSY preserving vacua and study the deviations
from such a SUSY limit after we taking into account the supergravity effects.
Consequently, constraints on the parameters $\epsilon\equiv \tl{\mu}/\Lambda_m$ from the lifetime of the meta-stable vacuum in ordinary ISS model will no longer be needed in our scenario.

After integrating out the mass terms $h\Phi$ of $\tl{q},q$, the low energy superpotential is
\beqa
\label{iss}
W_l=N_c(h^{N_F} \Lambda_m^{3N_c-N_F}\det\Phi)^{1/N_c}-h\tl{\mu}^2 {\rm Tr}(\Phi)~.
\eeqa
 The SUSY breaking effects from $F_\phi$ also prompts $F_\Phi$ to be nonzero at large values of $\Phi$.
 Adding the compensator field into the previous superpotential, we have
\beqa
W_l=N_c (h^{N_F} \Lambda_m^{3N_c-N_F}\det\Phi)^{1/N_c}\phi^{3-N_F/N_c}-h\tl{\mu}^2 \phi^2{\rm Tr}(\Phi)~,
\eeqa
where all fields within $\Phi$ have Weyl weight of 1.
When $F_\phi$ is turned on, the tree-level potential for the scalar $\Phi$ is
\beqa
V&=&|F_{\Phi_i^j}|^2-N_c(h^{N_F} \Lambda_m^{3N_c-N_F}\det\Phi)^{1/N_c}(3-\f{N_F}{N_c})F_\phi+2h\tl{\mu}^2F_\phi
 {\rm Tr}(\Phi)~,
\eeqa
which gives the minimum condition of $\langle\Phi\rangle\propto \tl{m} \delta_i^j$~
\beqa
2\(N_F \Lambda_m^{3-N_F/N_c} m^{N_F/N_c-1}-\tl{\mu}^2 N_F\)(N_F/N_c-1)N_F \Lambda_m^{3-N_F/N_c}m^{N_F/N_c-2} &&\nn\\
-N_F \Lambda_m^{3-N_F/N_c} m^{N_F/N_c-1}(3-\f{N_F}{N_c})F_\phi+2\tl{\mu}^2N_F F_\phi=0. &&
\eeqa
Here we use $m=h\tl{m}$. This equation is a transcendental equation which can not be solved exactly.
For a large $N_c$ with $N_c/N_F\ra 1$ (if the dual description is introduced as the input), we have
\beqa
m=\f{(N_F-N_c)\Lambda_m^2}{F_\phi}~,
\eeqa
which gives
\beqa
\f{F_\Phi}{\Phi}=-\f{h^2 N_F(\Lambda_m^2-\tl{\mu}^2)F_\phi}{(N_F-N_c)\Lambda_m^2}\approx -\f{h^2 N_F}{(N_F-N_c)}F_\phi,
\eeqa
when $\Lambda_m^2\gg\tl{\mu}^2$.
The low energy wave function only depends on $\tl{\Phi}=\Phi/\phi$ with
\beqa
\f{F_{\tl{\Phi}}}{\tl{\Phi}}=\f{F_\Phi}{\Phi}-F_\phi\approx -\f{h^2 N_F}{(N_F-N_c)}.
\eeqa
So we can see that we obtain a negatively deflected contribution.

In the limit of $N_F/N_c\ra 3/2$ which is $N_F\ra 3N$ in the original theory, we have
\beqa
\f{3}{2}\Lambda_m^{3/2}F_\phi (\sqrt{m})^2-(N_F\Lambda_m^3+2\tl{\mu}^2 F_\phi)\sqrt{m}+N_F \tl{\mu}^2 \Lambda_m^{3/2}=0.
\eeqa
A positive solution for $\sqrt{m}$ requires
\beqa
\Delta\equiv(N_F\Lambda_m^3+2\tl{\mu}^2 F_\phi)^2-6\Lambda_m^{3}F_\phi N_F \tl{\mu}^2 \geq0~.
\eeqa
So for $\Lambda_m\gg \tl{\mu},F_\phi$, we can obtain
\beqa
m\approx \f{4}{9} N_F^2\Lambda_m^3/F_\phi^2,~~~~~ m\approx \f{\tl{\mu}^4}{9\Lambda_m^{3}}.
\eeqa
Here only the first solution depends on $F_\phi$ and therefore we keep such a anomaly mediation
contribution solution.
Then we can obtain the deflection parameter
\beqa
\f{F_{\tl{\Phi}}}{\tl{\Phi}}=\f{F_\Phi}{\Phi}-F_\phi\approx -(\f{3}{2}h^2+1)F_\phi.
\eeqa
In this limit, the deflection parameter is still negative.

In the limit $N_F/N_c\ra 3$ with $N_F\lesssim 3N_c$
(which amounts to $N_F>3/2N$ and the theory lies in the conformal window), the '{\em magnetic}' theory will no longer be IR free.
 However, the dynamically superpotential will still have the form (\ref{iss}).
 Following our previous discussions, the resultant cubic equation takes the form
\beqa
2m^3-2\tl{\mu}^2 m+2\f{F_\phi}{N_F}=0,
\eeqa
in this limit which can always give negative solutions for $m$ due
to the continuous nature of the cubic function.
Numerical calculations indicate that the expression
\beqa
\f{F_\Phi}{\Phi}&=&-\f{h^2 N_F (m^2-\tl{\mu}^2)}{m}\equiv c h^2 F_\phi,\nn\\
 &\approx& 0.53 h^2 N_F\tl{\mu}\approx 0.26 h^2 F_\phi,~~~{\rm when}~ m\approx -1.3 \tl{\mu}, F_\phi= 2 N_F \tl{\mu},\nn\\
 &\approx& 0.04 h^2 N_F\tl{\mu}\approx 0.4 h^2 F_\phi,~~~~{\rm when}~ m\approx -1.02 \tl{\mu}, F_\phi= 0.1 N_F \tl{\mu},
\eeqa
always gives a coefficient $c$ less than $1$. Therefore we can obtain the deflection parameter for a
large $N_c$
\beqa
d\equiv\f{F_{\tl{\Phi}}}{\tl{\Phi}F_\phi}=\f{F_\Phi}{\Phi F_\phi}-1\approx c h^2-1
\eeqa
with $0<c<1$. Depending on the size of the coupling {\em h}, the deflection parameter
can be positive or negative. We can see that there is a smooth transition between the
positive and negative deflected regions by adjusting the value of {\em h}  for general
choices of $N_F$ and $N_c$.
A positive deflection corresponds to a relatively large coupling {\em h}.

The general results of the deflected anomaly mediation scenario are given by \cite{deflect,okada}
\beqa
\f{m_{\la_i}}{\al(\mu)}&=&\f{F_\phi}{2}\(\f{\pa}{\pa \ln\mu}-d\f{\pa}{\pa\ln|\Phi|}\)\al^{-1}(\mu,\Phi),\nn\\
m_i^2(\mu)&=&-\f{|F_\phi|^2}{4}\(\f{\pa}{\pa \ln\mu}-d\f{\pa}{\pa\ln|\Phi|}\)^2\ln Z_i(\mu,\Phi), \nn\\
A_i(\mu)&=&-\f{F_\phi }{2}\(\f{\pa}{\pa \ln\mu}-d\f{\pa}{\pa\ln|\Phi|}\)\ln Z_i(\mu,\Phi).
\eeqa
 The gaugino masses which acquire an additional contributions from gauge mediation are given by
\beqa
\label{gaugino}
m_{\la_i}(\mu)=\f{\al_i(\mu)}{4\pi}F_\phi(b_i+d N_F),
\eeqa
with $b_i$ being the beta functions for the gauge couplings.
Similarly, the contribution for sfermions are
\beqa
m_i^2(\mu)&=&2\sum\limits_{G_i}C_2(r)\(\f{\al(\mu)}{4\pi}\)^2|F_\phi|^2b_i G(\mu,\Phi)~,
\eeqa
with
\beqa
&&G(\mu,\Phi)=\(\f{N_F}{b}-\f{N_F^2}{b^2}\)\xi^2 d^2+\(\f{N_F }{b}d+1\)^2,\\
&&\xi\equiv\f{\al(\Phi)}{\al(\mu)}=\[1+\f{b}{4\pi}\al(\Phi)\ln\(\f{\Phi^\da \Phi}{\mu^2}\)\].
\eeqa
Here the MSSM beta function are $b_i=(-33/5,-1,3)$ and
the quadratic Casimir for $SU(N)$ fundamental representation is $C_2(N)=(N^2-1)/2N$.
The trilinear couplings $A_\la$ related to superpotential terms $\la_{ijk} Q_iQ_jQ_k$
are given by $A_\la=(A_{Q_i}+A_{Q_j}+A_{Q_k})\la_{ijk}$ with
\beqa
A_i(\mu)=2 c_i\[-\al(\mu)+\f{d N_F}{b}\(\al(\Phi)-\al(\mu)\)\] \f{F_\phi}{4\pi}+|y(\mu)|^2\f{F_\phi}{32\pi^2}.
\eeqa
The generations of higgsino mass $\mu$ and the soft SUSY breaking $B_\mu$ are not straight forward and
should be seen as an independent problem of anomaly mediation. For example,
they can be generated by the mechanism proposed in \cite{rattazzi} with additional singlet $S$.
Therefore, we will consider them as free parameters and do not give
their explicit expressions in terms of the model inputs.

\section{Reconcile $g_\mu-2$ and LHC data in deflected AMSB} \label{phenomenology}

The SM prediction of the muon anomalous magnetic moment is
\beqa
a^{\rm SM}_\mu =116591834(49)\times 10^{-11}~,
\eeqa
which is smaller than the experimental result of E821 at the Brookhaven AGS \cite{Mg-2:collaboration}
\beqa
a_\mu^{\rm expt} =116592089(63)\times 10^{-11}~.
\eeqa
 The deviation is then about $3\sigma$
 \beqa
\Delta a_\mu({\rm expt - SM}) = (255\pm 80)\times 10^{-11}.
\eeqa
SUSY can yield sizable contributions to the muon $g_\mu-2$ which dominately
come from the chargino-sneutrino and the neutralino-smuon loop diagrams.
At the leading order the analytic expressions for the SUSY contributions
are presented in \cite{Mg-2:Moroi}. The $g_\mu-2$ anomaly,
 which is order $10^{-9}$,  can be explained for $m_{\rm SUSY} = {\cal O}(100)$ GeV
and $\tan\beta = {\cal O}(10)$.

Now we turn to the calculations in the deflected AMSB.
From the expression of deflected AMSB spectrum,
the gaugino mass ratio can be enlarged in the positive deflected scenario while
diminished in the negative deflected case.
The gaugino mass ratio at the electroweak scale is given by
\beqa
M_3:M_2:M_1\approx  12:12:11.6\approx 1:1:1,
\eeqa
with $d=-1$ and $N_F=5$.
On the other hand, for the positively deflected AMSB scenario, we have at the electroweak scale
\beqa
M_3:M_2:M_1\approx  -48:-8:1.6 \approx -30:-5:1,
\eeqa
with $d=1$ and $N_F=5$. So we can see that the $g_\mu-2$ anomaly
may be explained in the positively deflected AMSB scenario \cite{okada2}.

We scan the parameter space of our deflected scenarios with
$3\geq d\geq -3$, $N_F\geq 5$ and the messenger scale $M$.
The messenger scale $M = \Phi$ plays a role of the intermediate
threshold between the UV cutoff and the electroweak scale.
At the messenger scale $M$, the gaugino soft masses are given by
\beqa
\label{gaugino}
m_{\la_i}(M)=\f{\al_i(M)}{4\pi}F_\phi(b_i+d N_F).
\eeqa
The sfermion masses at the messenger scale $M$ are
\beqa
\f{m_{\tl{Q}_L}^2}{|F_\phi|^2}&=&\f{\al^2_3(M)}{(4\pi)^2}8G_3-\f{\al^2_2(M)}{(4\pi)^2}\f{3}{2}G_2-\f{\al^2_1(M)}{(4\pi)^2}\f{11}{50}G_1,\\
\f{m_{\tl{U}^c_L}^2}{|F_\phi|^2}&=&\f{\al^2_3(M)}{(4\pi)^2}8G_3-\f{\al^2_1(M)}{(4\pi)^2}\f{88}{25}G_1,\\
\f{m_{\tl{D}^c_L}^2}{|F_\phi|^2}&=&\f{\al^2_3(M)}{(4\pi)^2}8G_3-\f{\al^2_1(M)}{(4\pi)^2}\f{22}{25}G_1,\\
\f{m_{\tl{L}_L}^2}{|F_\phi|^2}&=&-\f{\al^2_2(M)}{(4\pi)^2}\f{3}{2}G_2-\f{\al^2_1(M)}{(4\pi)^2}\f{99}{50}G_1,\\
\f{m_{\tl{E}_L^c}^2}{|F_\phi|^2}&=&-\f{\al^2_1(M)}{(4\pi)^2}\f{198}{25}G_1,
~~~\f{m_{\tl{H}_d}^2}{|F_\phi|^2}=\f{m_{\tl{L}_L}^2}{|F_\phi|^2}~,\\
\f{m_{\tl{H}_u}^2}{|F_\phi|^2}&=&\f{m_{\tl{L}_L}^2}{|F_\phi|^2}-3\f{y_t^2}{(16\pi^2)^2}(\f{16}{3}g_3^2+3g_2^2+\f{13}{15}g_1^2-6y_t^2),
\eeqa
where we define
\beqa
G_i&=&\(\f{N_F}{b_i}-\f{N_F^2}{b_i^2}\) d^2+\(\f{N_F }{b_i}d+1\)^2.
\eeqa
The stop soft masses should also include the yukawa contributions
\beqa
\f{m_{\tl{Q}_{L,3}}^2}{|F_\phi|^2}&=&\f{m_{\tl{Q}_L}^2}{|F_\phi|^2}-\f{y_t^2}{(16\pi^2)^2}(\f{16}{3}g_3^2+3g_2^2+\f{13}{15}g_1^2-6y_t^2)~,\\
\f{m_{\tl{t}_{L}^c}^2}{|F_\phi|^2}&=&\f{m_{\tl{U}^c_L}^2}{|F_\phi|^2}-2\f{y_t^2}{(16\pi^2)^2}(\f{16}{3}g_3^2+3g_2^2+\f{13}{15}g_1^2-6y_t^2)~.
\eeqa
The trilinear soft terms are given by
\beqa
\f{A_t}{(F_\phi/2\pi)} &=&-\f{8}{3}\al_3(M)-\f{3}{2}\al_2(M)-\f{13}{30}\al_1(M)+\f{1}{8\pi}\(6|y_t(M)|^2+{|y_b(M)|^2}\),\\
\f{A_b}{(F_\phi/2\pi)} &=&-\f{8}{3}\al_3(M)-\f{3}{2}\al_2(M)-\f{7}{30}\al_1(M)\nn\\
                      && +\f{1}{8\pi}\(|y_t(M)|^2+6{|y_b(M)|^2}+|y_{\tau}(M)|^2\),\\
\f{A_\tau}{(F_\phi/2\pi)} &=&-\f{3}{2}\al_2(M)-\f{9}{10}\al_1(M)+\f{1}{8\pi}\(3{|y_b(M)|^2}+4|y_{\tau}(M)|^2\).
\eeqa
Note that here we use the notation $g_Y^2= 3g_1^2 /5$.
The tachyonic slepton problem can be solved with the choice of $d$ and $N_F$.

The inputs should be seen as the boundary conditions at the messenger scale which,
after RGE running to the electroweak scale, should give the low energy spectrum.
The free parameters are chosen as $d,N_F,F_\phi,M,\tan\beta$.
We scan over the following ranges of these parameters:
\bit
\item In our scenario, the value of $F_\phi$ determines the whole spectrum.
Constraints from the gaugino masses indicate that $F_\phi$ cannot be too low.
Thus, we choose $F_\phi\gtrsim {\cal O}(10 {\rm TeV})$. On the other hand,
a too heavy $F_\phi$ will spoil the EWSB and lead to too heavy Higgs mass.
In our scan we take the value of $F_\phi$ in the range $10 {\rm TeV}<F_\phi<500 {\rm TeV}$.
\item The messenger scale $M$ can be chosen to be below the typical GUT scale $10^{16} {\rm GeV}$.
It should be heavier than the sparticle spectrum.
The lower bound is chosen to be ${\cal O}(10 {\rm TeV})$.
We note that possible Landau pole problem can possibly be avoided by setting the dynamical scale of
the ISS sector to be high enough in a way that is compatible with phenomenological requirements.
\item We choose $N_F\geq5$ and $3\geq d\geq -3$. The value of $\tan\beta$ is chosen to be
$40\geq\tan\beta\geq 2$.
\item  The parameter $\mu$ is chosen to have the same sign as $M_2$ because this case
gives positive SUSY contributions to $g_\mu-2$.
\eit
In our scan we take into account the following collider and dark matter constraints:
\bit
\item[(1)] The lower bounds of LEP on neutralino and charginos masses, including the
invisible decay of $Z$-boson.
We require $m_{\tl{\chi}^\pm}> 103 {\rm GeV}$ and the invisible decay
width $\Gamma(Z\ra \tl{\chi}_0\tl{\chi}_0)<1.71~{\rm MeV}$,
which is consistent with the $2\sigma$ precision electroweak measurement
$\Gamma^{non-SM}_{inv}< 2.0~{\rm MeV}$.
\item[(2)] For the precision  electroweak measurements,
we require the oblique '{\em S,T,U}' parameters \cite{oblique} to be
be compatible with the LEP/SLD data at 2$\sigma$ level \cite{stuconstraints}.
\item[(3)] The combined mass range for the Higgs boson: $123 {\rm GeV}<M_h <127 {\rm GeV}$
from ATLAS and CMS \cite{ATLAS:higgs,CMS:higgs}.
\item[(4)] The relic density of the neutralino dark matter satisfies the Planck result
$\Omega_{DM} = 0.1199\pm 0.0027$ \cite{planck} (in combination with the WMAP data \cite{wmap}).
\eit

\begin{figure}[htbp]
  \begin{minipage}[t]{0.5\linewidth}
    \centering
    \includegraphics[width=3in]{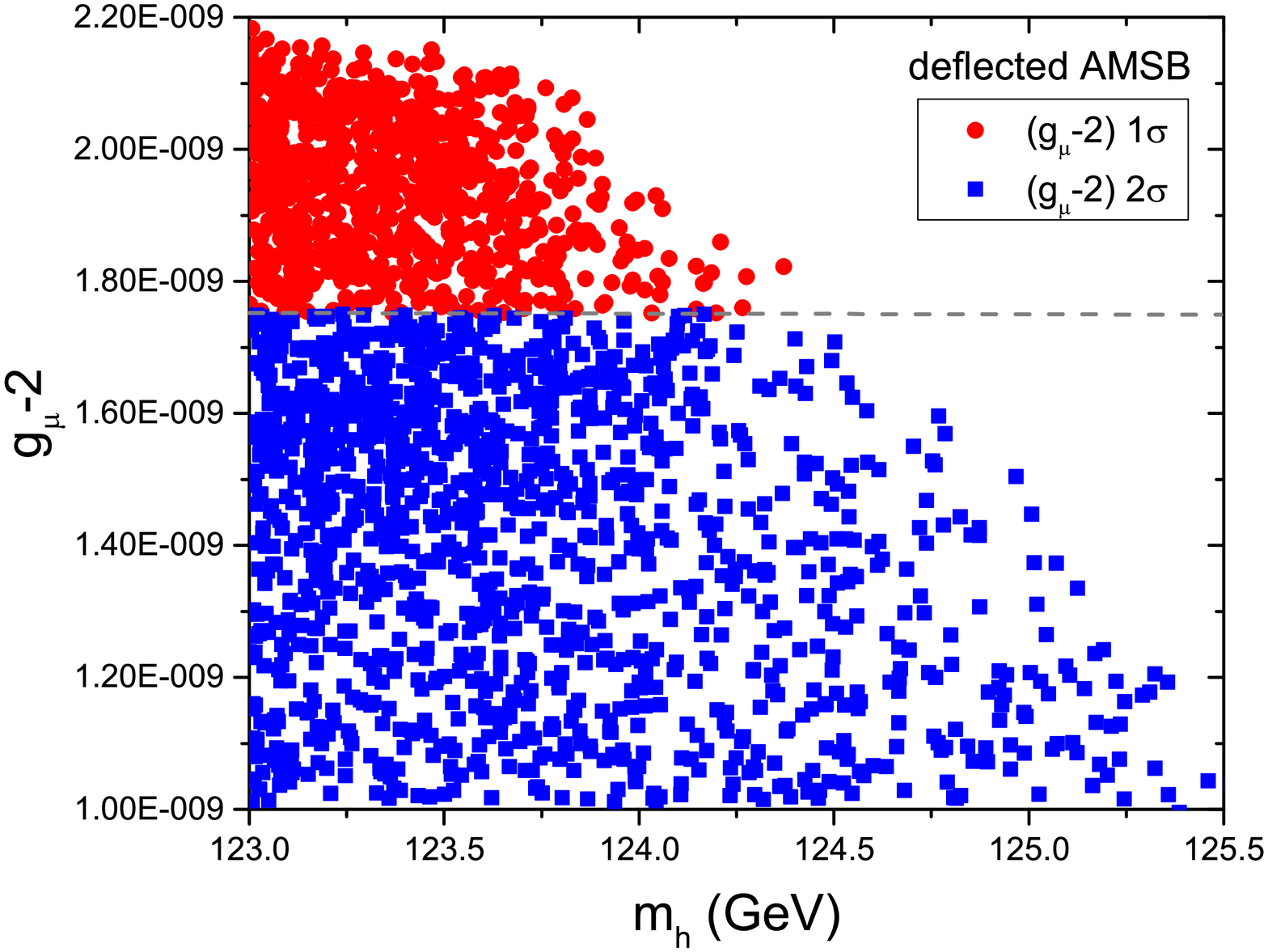}
  \end{minipage}
  \begin{minipage}[t]{0.5\linewidth}
    \centering
    \includegraphics[width=3in]{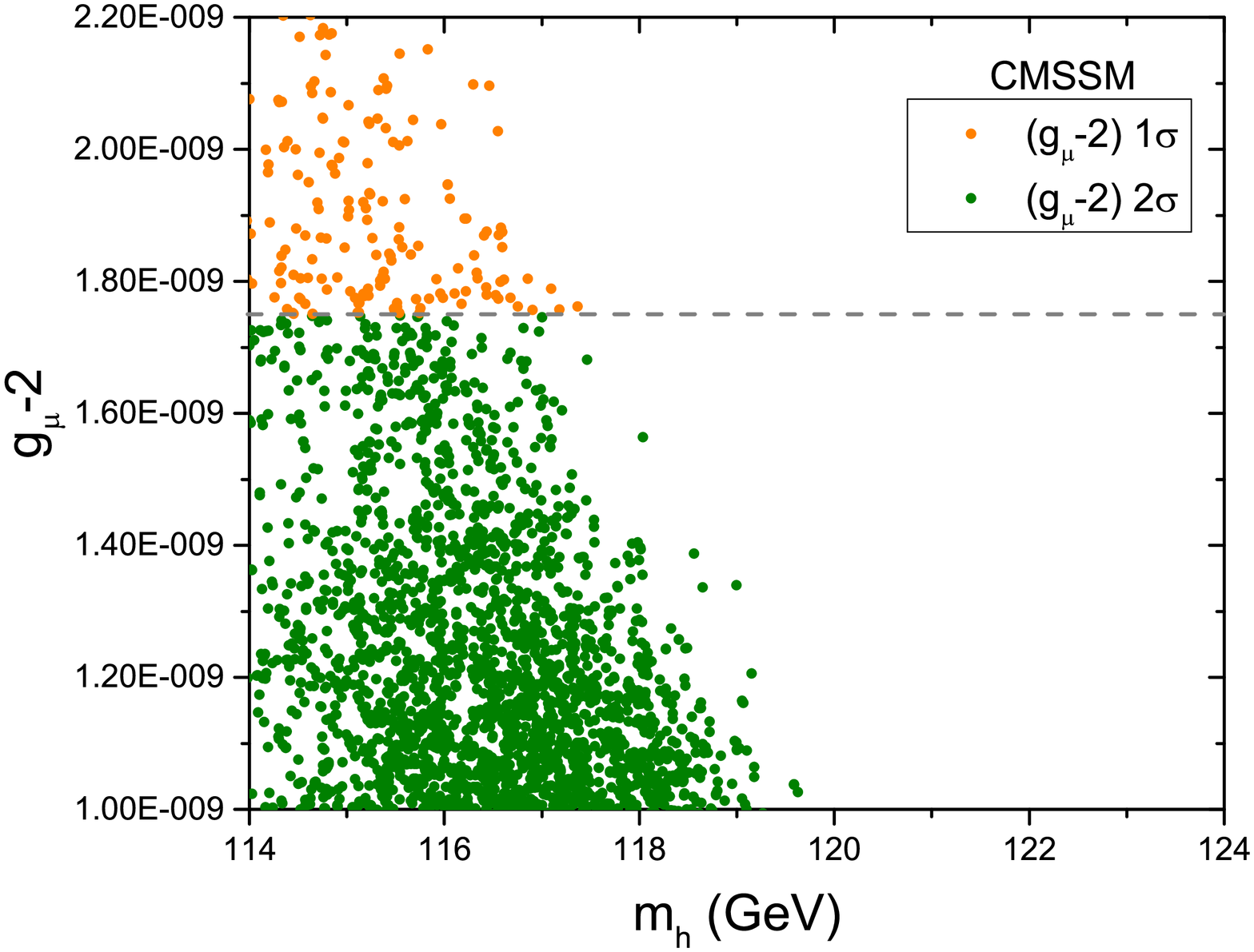}
  \end{minipage}
\vspace*{-1.0cm}
\caption{The left and right panels show the scatter plots of the parameter space for
our deflected AMSB scenario and the CMSSM, respectively.
All the points survive the collider and dark matter
constraints (1-4). }
\label{fig1}
\end{figure}

\begin{figure}[htbp]
  \begin{minipage}[t]{0.5\linewidth}
    \centering
    \includegraphics[width=3in]{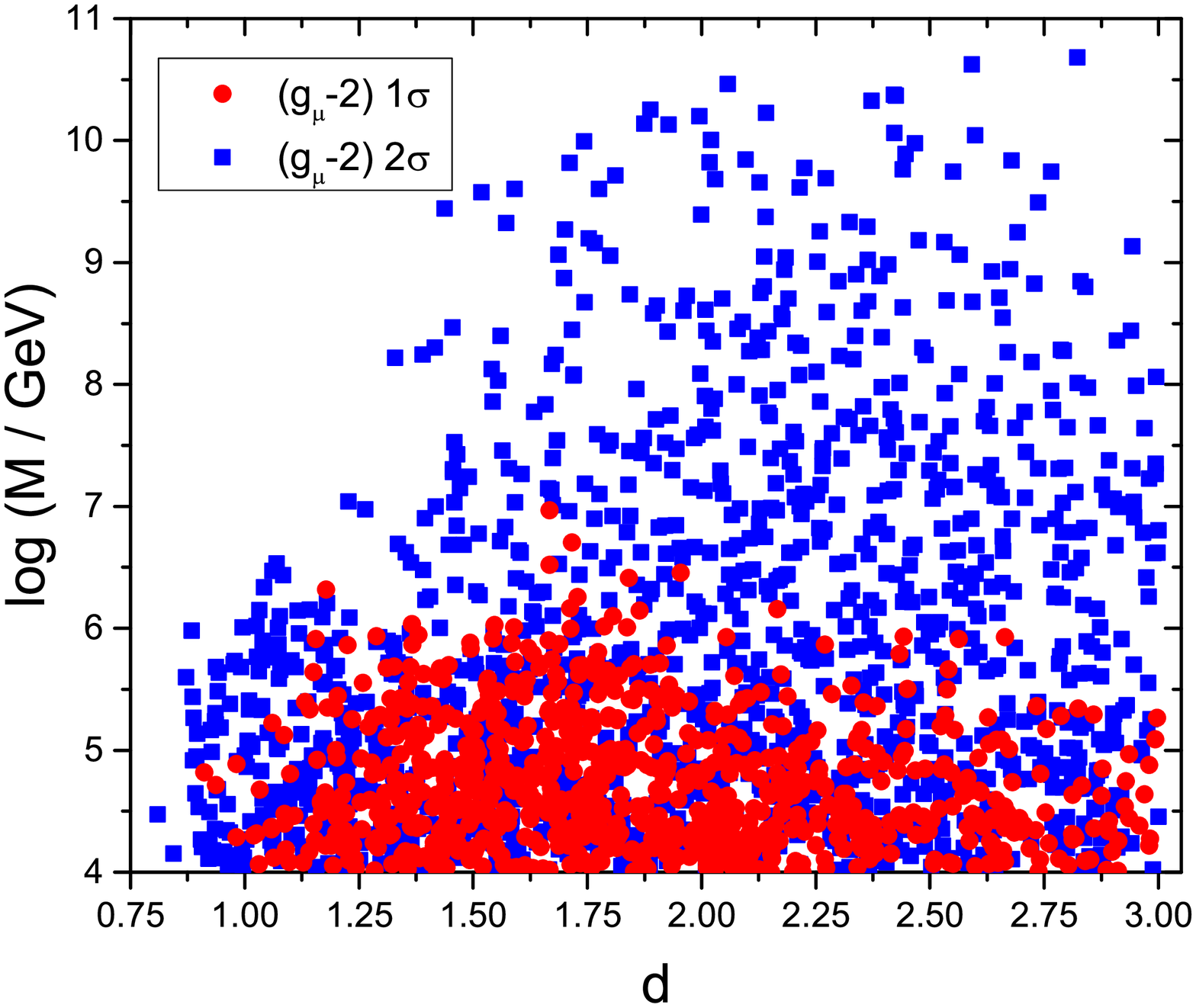}
  \end{minipage}
  \begin{minipage}[t]{0.5\linewidth}
    \centering
    \includegraphics[width=3in]{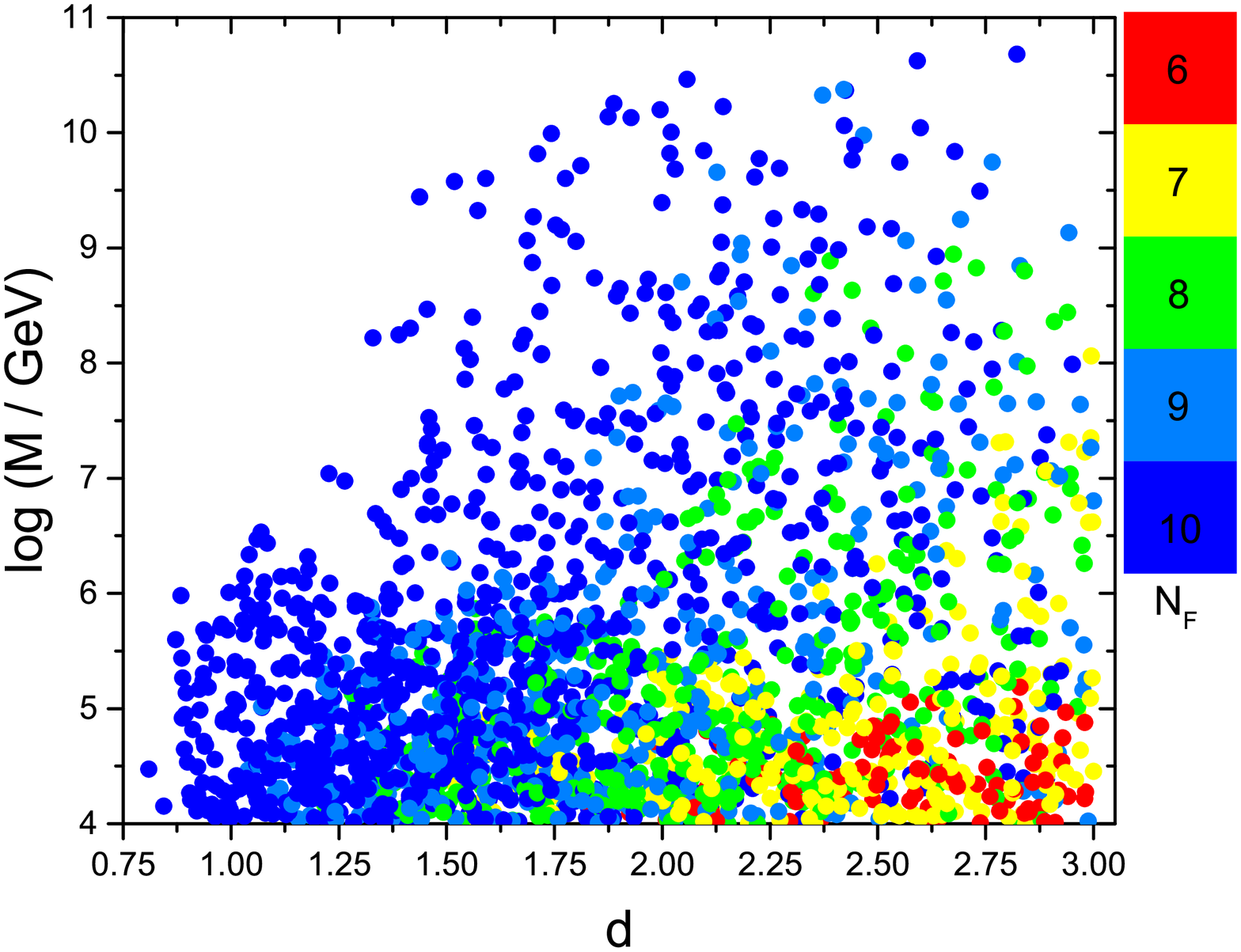}
  \end{minipage}
\vspace*{-1.0cm}
\caption{Same as Fig.\ref{fig1}, but showing the deflection parameter $d$ versus
the messenger scale $M$ for our deflected AMSB scenario.}
\label{fig2}
\end{figure}

\begin{figure}[htbp]
  \begin{minipage}[t]{0.5\linewidth}
    \centering
    \includegraphics[width=3in]{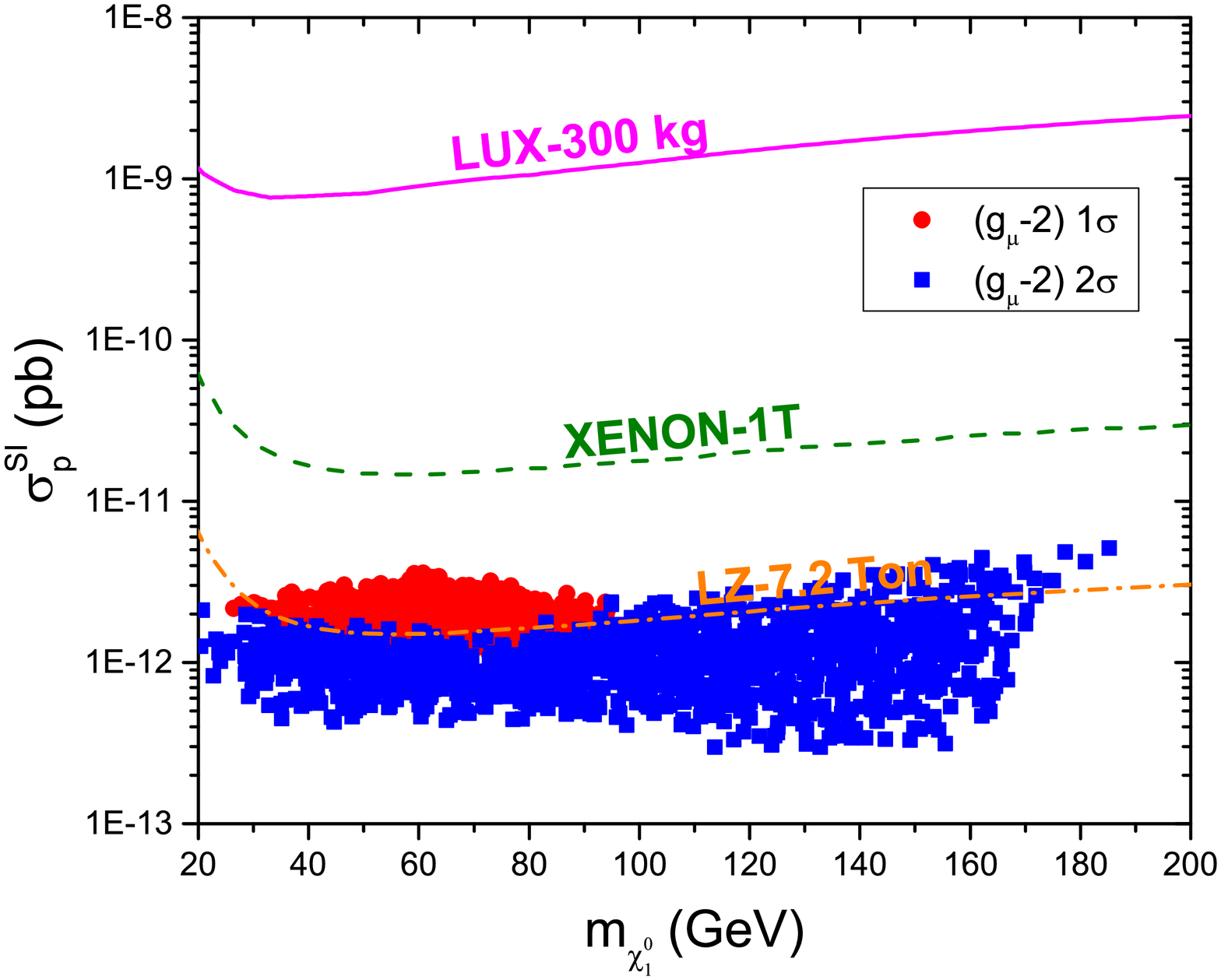}
  \end{minipage}
  \begin{minipage}[t]{0.5\linewidth}
    \centering
    \includegraphics[width=3in]{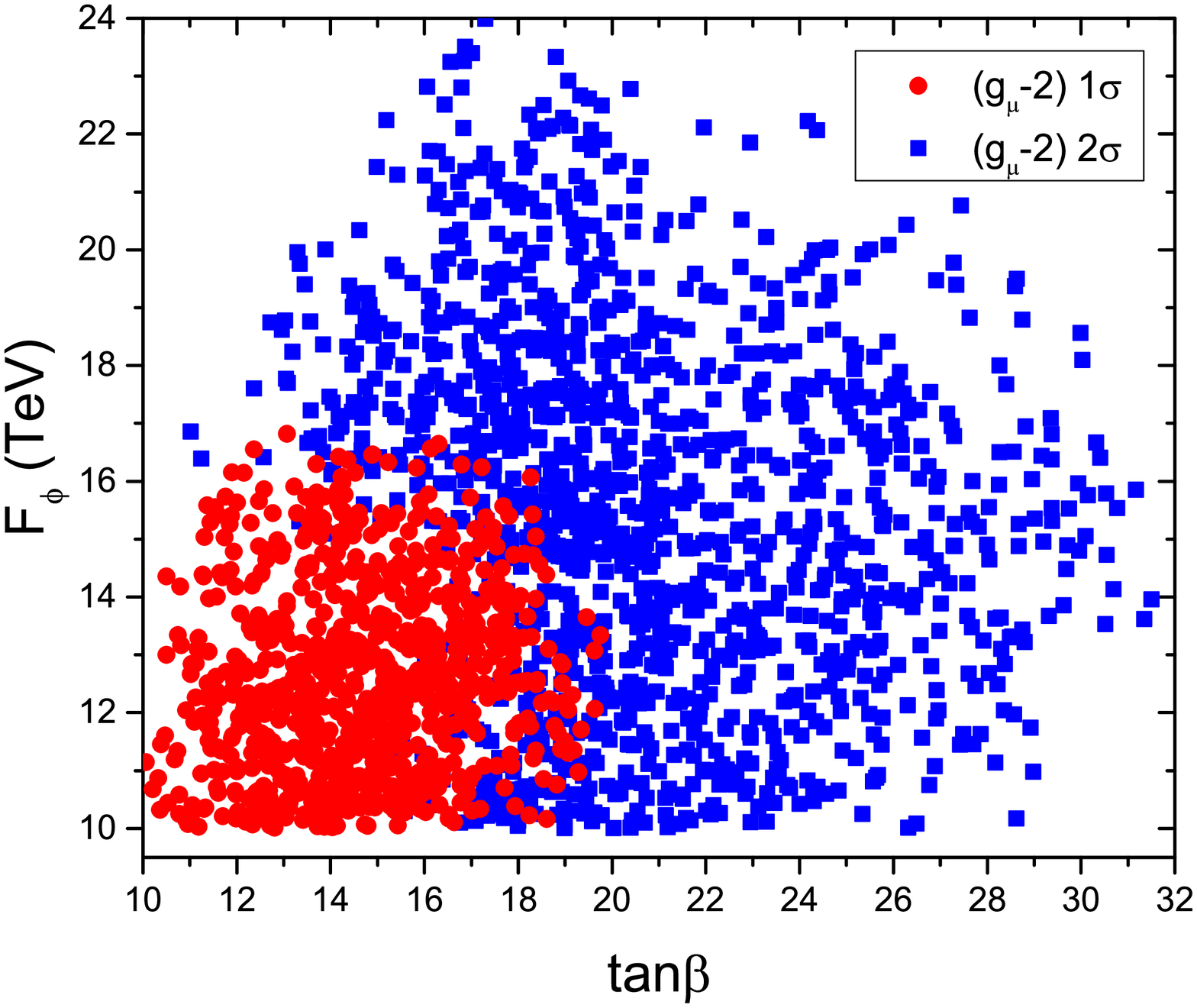}
  \end{minipage}
\vspace*{-1.0cm}
 \caption{Same as Fig.\ref{fig1}, but showing the spin-independent cross section of dark matter (the
lightest neutralino) scattering off the nucleon versus the dark matter mass in the left panel
and $F_\Phi$ versus $\tan\beta$ in the right panel.
The LUX \cite{lux} limits and the XENON1T \cite{xenon} and LUX-ZEPLIN 7.2 Ton \cite{LZ-72}
sensitivities are plotted.}
\label{fig3}
\end{figure}

In Figs.\ref{fig1}-\ref{fig3}, we show the samples that
survive the above constraints. From these figures we obtain the following observations:
\bit
\item[(i)]
Our scenario can account for both the $g_\mu-2$ anomaly and current Higgs mass measurement at the LHC.
It is clear from Fig.\ref{fig1} that in order to solve the $g_\mu-2$  anomaly at 2$\sigma$ level,
the Higgs mass can reach 125.5 GeV (see the blue points in the left panel).
However, to solve the $g_\mu-2$ anomaly at 1$\sigma$ level, the Higgs boson mass
cannot be in the best range $125.09\pm0.24$ GeV \cite{higgs:latest}
(the red points in the left panel are upper bounded by 124.5 GeV).
Such results are not surprising because the stop sector can give sizeable contributions to
Higgs mass only if $A_t$ is large enough which however is controlled by $F_\phi$ in our scenario.
As shown in the right panel of Fig.\ref{fig1},
in the CMSSM the Higgs boson mass is upper bounded by 120 GeV in order to
solve the $g_\mu-2$ anomaly at 2$\sigma$ level. So, our scenario is much better in
solving the $g_\mu-2$ anomaly and satisfying the Higgs mass measurement.

\item[(ii)] From Fig.\ref{fig2} we see the relations between the
deflection parameter $d$, the messenger mass scale $M$ and $N_F$.
We find that $M$ is constrained to below $10^{11}$ GeV if the $g_\mu-2$ anomaly
is solved at 2$\sigma$ level. This upper bound on $M$ is lowered to $10^7$ GeV
if the $g_\mu-2$ anomaly is solved at 1$\sigma$ level.
A low $N_F$ value corresponds to a relatively high deflection parameter $d$.
Besides, $d$ is constrained in the range $0.7<d<3$.
 It is clear from the right panel of Fig.\ref{fig3} that the value of $\tan\beta$ lies in the
range $10<\tan\beta< 20$ in order to explain the muon $g_\mu-2$ discrepancy at 1$\sigma$
level. The value of $F_\phi$ which determines the whole sparticle spectrum is upper
bounded by 17 TeV (25 TeV) at  1$\sigma$ (2$\sigma$) level.

\item[(iii)] In the ordinary AMSB scenario, dark matter is mostly wino like
which should be 2.7-3 TeV to provide enough cosmic dark matter content.
Since the direct detection cross section for the pure wino
is extremely small, below ${\cal O}(10^{-47}{\rm  cm^2})$, it is very difficult
to discover such a wino dark matter via direct detections.
In our positively deflected AMSB scenario with $N_F\geq 4$ and $d\geq 1$,
the lightest gaugino will in general no longer be wino. We can see from
the gaugino input that wino is always heavier than bino with large $N_F$ and
positive $d\sim {\cal O}(1)$. So the dark matter in our scenario
can be either bino-like or higgsino-like. In this case,
the dark matter may be accessible at the direct detections.
As shown in the left panel of Fig.\ref{fig3},
the parameter space for explaining the muon $g_\mu-2$ discrepancy at 1$\sigma$
level can be fully covered by the future LUX-ZEPLIN 7.2 Ton experiment  \cite{LZ-72}.
\eit

Finally, we show the details for two benchmark points in Table 1 and Table 2. The benchmark
points shown in these tables have $d>0$ and  $d<0$, respectively.

\begin{table}[h]\caption{A benchmark point with $d>0$. All the quantities with mass dimension are in GeV.}
\centering
\begin{tabular}{|c|c|c|c|c|}
\hline
$N_F$     & d         & M           & $F_\phi$           & $tan\beta$ \\ \hline
10        & 1.59      & $1.09\times 10^4$    & $1.33\times 10^4$    & 15.0       \\ \hline
\hline
$m_{\tilde{H}_u}^2$   & $m_{\tilde{H}_d}^2$   & $M_1$                 & $M_2$             & $M_3$             \\ \hline
$6.98\times 10^4 $             & $1.20\times 10^5 $             & $1.82\times10^2$              & $5.48\times 10^2$          & $1.88\times 10^3$          \\ \hline
$m_{\tilde{Q}_{L}}$   & $m_{\tilde{U}_{L}}$   & $m_{\tilde{D}_L}$     & $m_{\tilde{L}_L}$ & $m_{\tilde{E}_L}$ \\ \hline
$1.30\tm 10^3$              & $1.26\tm 10^3$              & $1.26\tm 10^3$              & $3.46\tm10^2$          & $1.53\tm10^2$          \\ \hline
$m_{\tilde{Q}_{L,3}}$ & $m_{\tilde{U}_{L,3}}$ & $m_{\tilde{D}_{L,3}}$ & $A_U$             & $A_D$             \\ \hline
$1.30\tm10^3$              & $1.25\tm10^3$              & $1.26\tm10^3$              & $-6.58\tm10^2$         & $-6.50\tm10^2$         \\ \hline
$A_L$                 & $A_\tau$              & $A_t$                 & $A_b$             &                   \\ \hline
$-1.46\tm10^2$             & $-1.17\tm10^2$             &$ -2.28\tm10^2 $            &$ -5.34\tm10^2 $        &                   \\ \hline
\hline
$Br(B\rightarrow X_S\gamma)$&$Br(B_S^0\rightarrow \mu^+\mu^-)$& $g_\mu-2$           & $\Omega_\chi h^2$          & $\sigma_P^{SI}$  \\ \hline
$3.25\tm10^{-4}$                    & $3.40\tm 10^{-9}$                        & $1.82\tm10^{-9}$           & 0.117                      & $1.09\tm10^{-12}$ pb      \\ \hline
$m_{h_1}$                   & $m_{\tilde{\chi}_1^0}$          & $m_{\tilde{\tau}_1}$& $m_{\tilde{\chi}_1^{\pm}}$ & $m_{\tilde{g}}$  \\ \hline
124.4                       & 84.1                            & 100.2               & 464.5                      & 3949.4           \\ \hline
\end{tabular}
\end{table}

\begin{table}[h]\caption{A benchmark point with $d<0$. All the quantities with mass dimension are in GeV.
In this case the LSP is $\tilde{\tau}$ and thus $\Omega_\chi h^2$ and $\sigma_P^{SI}$  can not be calculated.}
\centering
\begin{tabular}{|c|c|c|c|c|}
\hline
$N_F$     & d         & M           & $F_\phi$           & $tan\beta$ \\ \hline
10        & -2.66     & $4.57\tm10^6$    & $1.83\tm10^4$    & 12.2       \\ \hline
\hline
$m_{\tilde{H}_u}^2$   & $m_{\tilde{H}_d}^2$   & $M_1$                 & $M_2$             & $M_3$             \\ \hline
$4.78\tm10^3 $             & $7.12\tm10^4$              & $-1.02\tm10^3$             &$ -1.44\tm10^3 $        & $-2.54\tm10^3 $        \\ \hline
$m_{\tilde{Q}_{L}}$   & $m_{\tilde{U}_{L}}$   & $m_{\tilde{D}_L}$     & $m_{\tilde{L}_L}$ & $m_{\tilde{E}_L}$ \\ \hline
$8.40\tm10^2$              & $8.01\tm10^2 $             & $7.99\tm10^2 $             & $2.67\tm10^2 $         &$ 1.11\tm10^2$         \\ \hline
$m_{\tilde{Q}_{L,3}}$ & $m_{\tilde{U}_{L,3}}$ & $m_{\tilde{D}_{L,3}}$ & $A_U$             & $A_D$             \\ \hline
$8.27\tm10^2$              & $7.73\tm10^2$              & $7.99\tm10^2$              & $-7.57\tm10^2$         & $-7.45\tm10^2$         \\ \hline
$A_L$                 & $A_\tau$              & $A_t$                 & $A_b$             &                   \\ \hline
$-2.12\tm10^2$             & $-1.93\tm10^2$             & $-2.73\tm10^2$             & $-6.38\tm10^2$         &                   \\ \hline
\hline
$Br(B\rightarrow X_S\gamma)$&$Br(B_S^0\rightarrow \mu^+\mu^-)$& $g_\mu-2$           & $\Omega_\chi h^2$          & $\sigma_P^{SI}$  \\ \hline
$3.27\tm 10^{-4}$                    & $3.38\tm10^{-9} $                       & $-2.0\tm10^{-10} $             & -                          & -     \\ \hline
$m_{h_1}$                   & $m_{\tilde{\chi}_1^0}$          & $m_{\tilde{\tau}_1}$& $m_{\tilde{\chi}_1^{\pm}}$ & $m_{\tilde{g}}$  \\ \hline
125.6                       & 476.8                           & 383.5               & 1231.4                     & 5229.1           \\ \hline
\end{tabular}
\end{table}

\section{Conclusion}\label{conclusion}
We proposed a deflected anomaly mediation scenario from SUSY QCD
which can lead to both positive and negative deflection parameters.
There is a smoothly transition between these two deflection parameter regions
by adjusting certain couplings. This scenario can naturally
have a SUSY spectrum with heavy colored sparticles and light sleptons.
The discrepancy between the Brookheaven $g_\mu-2$ experiment and the LHC data can be reconciled.
We also found that the parameter space for explaining the muon $g_\mu-2$ discrepancy at 1$\sigma$
level can be fully covered by the future LUX-ZEPLIN 7.2 Ton experiment.

\section*{Acknowledgement}
This work was supported by the
Natural Science Foundation of China under grant numbers 11105124, 11105125, 11375001, 11172008,
11275245, 11135003,
the Innovation Talent project of Henan Province under grant number 15HASTIT017,
the Young-Talent Foundation of Zhengzhou University, the Ri-Xin Foundation of BJUT,
and Youth-Talents Foundation of eduction department of Beijing,
and by the CAS Center for Excellence in Particle Physics (CCEPP).

\end{document}